\documentclass[conference]{IEEEtran}
\IEEEoverridecommandlockouts
\IEEEpubid{\makebox[\columnwidth]{ 979-8-3315-8342-2/26/\$31.00 ~\copyright2026 IEEE \hfill} \hspace{\columnsep}\makebox[\columnwidth]{ }}

\usepackage{cite}
\usepackage{amsmath,amssymb,amsfonts}
\usepackage{algorithmic}
\usepackage{graphicx}
\usepackage{textcomp}
\usepackage{xcolor}
\usepackage{xspace}
\usepackage{bm}
\usepackage{comment}
\usepackage{caption}
\usepackage{subcaption}
\usepackage{hyperref}

\def\BibTeX{{\rm B\kern-.05em{\sc i\kern-.025em b}\kern-.08em
    T\kern-.1667em\lower.7ex\hbox{E}\kern-.125emX}}
\begin{document}

\title{Play and Learn: Gamified Feedback for Ultrasound-Guided Catheter Insertion Training in Virtual Reality\\
}



\author{
\IEEEauthorblockN{
Mohammad Raihanul Bashar\IEEEauthorrefmark{1},
Alejandro Olivares Hernandez\IEEEauthorrefmark{2},
Yahia Zine\IEEEauthorrefmark{2},
Anil Ufuk Batmaz\IEEEauthorrefmark{1}
}

\IEEEauthorblockA{\IEEEauthorrefmark{1}
Department of Computer Science \& Software Engineering\\
Concordia University, Montreal, Canada\\
Email: \{mohammadraihanul.bashar, ufuk.batmaz\}@concordia.ca}

\IEEEauthorblockA{\IEEEauthorrefmark{2}
Sononurse VS Inc., Montreal, Canada\\
Email: \{alejandro.olivares, yahia.zine\}@sononurse.com}
}

\maketitle
\IEEEpubidadjcol

\begin{abstract}
VR is widely used in procedural medical training, but many simulators prioritize realism over formative feedback. We developed a gamified VR simulator for ultrasound-guided peripheral IV catheter insertion, using aligned visual and auditory feedback such as progress indicators, guidance, and rewards. In studies with novices (N=24) and clinicians (N=12), gamification reduced task time, improved usability, and lowered workload. Qualitative results showed clearer goals and greater confidence for novices, and improved pacing for experts. Overall, gamification can provide effective formative feedback in VR medical training.
\end{abstract}


\section{Introduction}

Peripheral intravenous catheter (PIVC) insertion is widely used for rapid intravenous access, with over two billion catheters used annually worldwide~\cite{alexandrou2018use}. However, first-attempt failure rates can reach 39\%, and catheter failure affects 30–50\% of patients due to complications such as infiltration, phlebitis, occlusion, dislodgement, and hematoma~\cite{indarwati2020incidence, skulec2020two}. Landmark-based techniques rely on visual and tactile assessment of superficial veins~\cite{piredda2017factors}, motivating ultrasound guidance, which improves cannulation success and reduces complications~\cite{alvarez2024ultrasound}. Yet ultrasound-guided PIVC introduces perceptual–motor challenges, requiring bimanual coordination, needle-plane alignment, and interpretation of dynamic 2D images for 3D spatial judgments~\cite{leibowitz2020ultrasound}. Novices often struggle with probe alignment, depth perception, and needle visualization, while conventional training provides limited feedback and subsurface anatomical insight~\cite{mckinney2023standardized, hemingway2024ultrasound}.

Virtual reality (VR) enables risk-free, repeatable procedural training with controlled anatomy, hidden anatomy visualization, and real-time feedback~\cite{andersen2021teaching, liu2023effectiveness, chang2024impact}. However, many VR simulators prioritize procedural realism over learning progression, motivation, and error correction.

We present a gamified VR system for ultrasound-guided PIVC training that supports probe–needle coordination, venipuncture, and spatial alignment through real-time simulated ultrasound. Task-aligned gamified feedback, including alignment guidance, progress indicators, and performance summaries, enhances engagement and learning while preserving procedural fidelity. Controlled studies with novices and clinicians show improved efficiency, reduced workload, and better user experience. Our contributions are: (1) a VR training system integrating real-time ultrasound simulation with task-aligned gamified feedback, and (2) empirical evidence of its effectiveness across expertise levels.

\section{Related Works}

\subsection{Training in VR}

VR is increasingly used in instructional, professional, and high-stakes training due to its ability to reduce cost, support experiential learning, and enable safe, repeatable practice~\cite{regal2022vr, xie2021review, fujiwara2024virtual, long2025application}. Prior work shows that VR can improve skill acquisition, retention, and transfer by providing controlled, immersive environments for complex tasks that are difficult or unsafe to replicate in the real world~\cite{lawson2016future, nguyen2021stress, murtinger2021cbrne, grabowski2015virtual, regal2022challenges}. Its multisensory nature supports active participation, diverse learning styles, and adaptive feedback, making it especially promising for healthcare training~\cite{uhl2022threat, alwashmi2024enhancing, marougkas2025adaptive, maddalon2024exploring, regal2022marcus, nasri2025towards, lindner2025knowledge}. Existing VR medical applications include procedural training, such as surgery, resuscitation, and dialysis~\cite{harrington2018development, jaskiewicz2019applicability, papagiannakis2020mages, chheang2021collaborative, al2021healthcare, bucher2019vreanimate, rettinger2021vr}, as well as conceptual learning in diagnosis and anatomy~\cite{makled2019pathogenius, izard2016virtual}. However, VR medical education remains early-stage and requires stronger empirical and pedagogical validation.

\subsection{Gamification}

Gamification uses game design elements in non-game contexts to increase motivation, engagement, and retention~\cite{hamari2014does, deterding2011game}. It has been framed as designing for gamefulness through challenge, fantasy, and curiosity~\cite{deterding2014eudaimonic}, or as adding motivational affordances to promote behavioral outcomes~\cite{huotari2012defining, ryan2006motivational}. Common elements such as points, badges, leaderboards, and progress indicators can support competence, feedback, and goal pursuit~\cite{ryan2006motivational, mekler2013points, landers2017gamification, hamari2017badges, cruz2017need}. Research shows that gamification can improve productivity and engagement across education, healthcare, and business, although outcomes vary by context and user characteristics~\cite{hamari2014does, fang2021hci, johnson2016gamification, zichermann2020gamification, deterding2011game}. Beyond traditional PBL mechanics~\cite{deterding2011game, hamari2014does, seaborn2015gamification}, critiques of superficial gamification have motivated richer social, affective, and audiovisual strategies~\cite{deterding2014eudaimonic, seaborn2015gamification, thibault2021seven}. In particular, \textit{juicy design} uses enhanced audiovisual feedback to increase engagement and game feel~\cite{gabler2005prototype, hicks2019juicy}. Accordingly, we combine PBL-style performance feedback with juicy design to gamify an ultrasound-guided VR catheter-insertion simulation.

\subsection{Effects of Gamification on Learning}

In education, gamification combines game elements, active learning, and motivational psychology to improve engagement, interest, self-efficacy, focus, and learning outcomes~\cite{deterding2011game, huotari2012defining, hamari2014does, karimov2024gamification, li2024gamification}. Although its effectiveness varies and it should complement rather than replace pedagogy, gamification is widely applied across educational contexts~\cite{putra2025gamification, diaz2024meta, lampropoulos2024impact, furkan2025meta}.

\section{VR Simulation of UG PIVC Insertion}\label{sec:design}

\begin{figure}[hbt]
    \centering
    \includegraphics[height=0.5\linewidth]{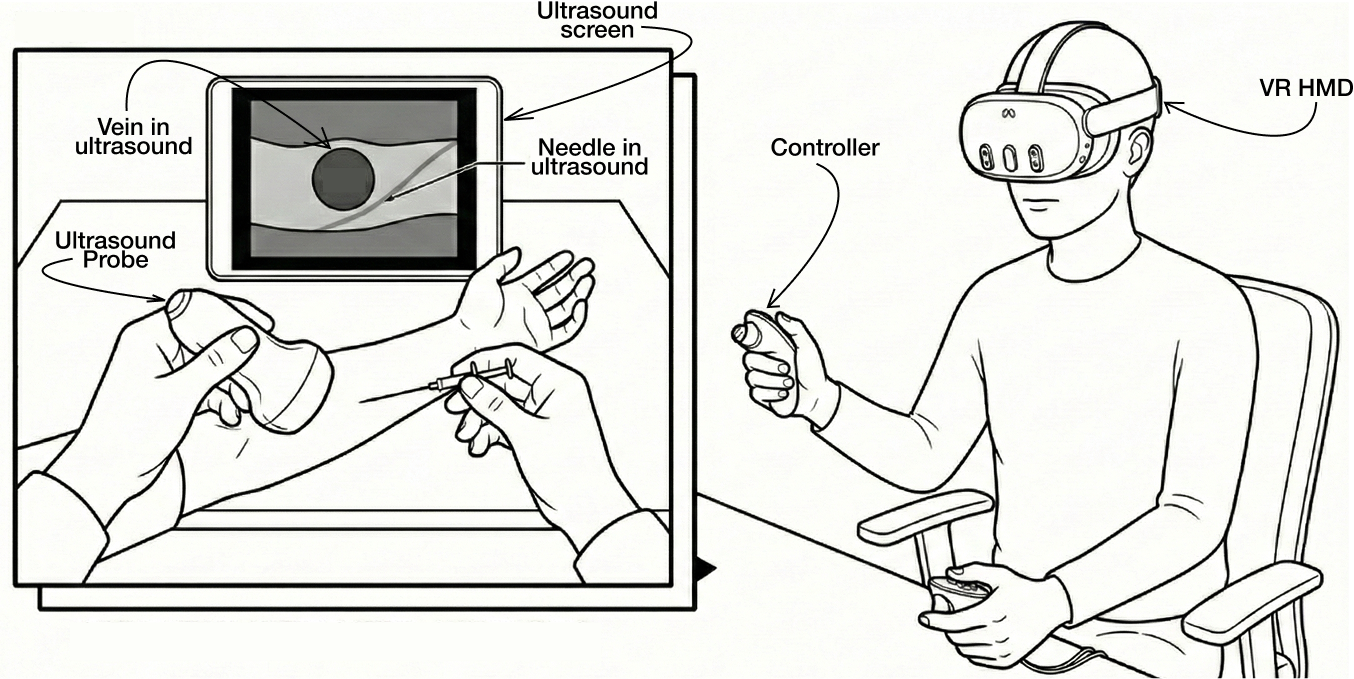}
    \caption{Technical setup of the VR ultrasound-guided PIVC insertion simulation.}
    \label{fig:system}
\end{figure}

We designed a VR simulation that translates the perceptual--motor demands of ultrasound-guided PIVC insertion into a controlled, repeatable training environment emphasizing real-time needle visualization, probe--needle coordination, and spatial alignment. Because continuous needle-tip visibility is critical for safe cannulation, the system focuses on the \textbf{In-Plane (Longitudinal)} technique, where the probe is aligned parallel to the vein to maintain full visualization of the vessel and advancing needle (see~\autoref{fig:system}). Although the system supports Out-of-Plane scanning for vein localization, the In-Plane approach is prioritized because it provides clearer feedback and is better suited to novice training.

\subsection{Technical Setup}

The simulation was built in \textit{Unity 2022.3} using the \textit{PlusToolkit Ultrasound Simulator} and ran on a high-performance PC with a Meta Quest 3 headset via Quest Link. Real-time ultrasound images were generated through OpenIGTLink streaming, while the anatomically accurate forearm model ran at 90 Hz for smooth interaction. The virtual environment depicted a medical classroom with a hospital bed, full-body mannequin, detailed forearm, ultrasound display, and visible virtual hands. Participants controlled the probe with the non-dominant hand and the needle with the dominant hand using VR controllers. The system provided real-time visual and auditory feedback but did not model haptic resistance due to the lack of force feedback in standard VR controllers.

\subsubsection{Simulated Catheter Insertion Process}\label{sec}

The simulation reproduced key stages of ultrasound-guided PIVC insertion: vein localization, probe alignment, needle advancement, and cannulation confirmation. Participants first scanned the forearm in the Out-of-Plane view to locate the vein, then rotated the probe into the In-Plane orientation for longitudinal visualization and continuous needle tracking. Successful vein entry triggered ultrasound deformation and a brief confirmation sound. After each success, probe and needle interactions were disabled for five seconds and then reset to maintain consistent pacing.

\subsubsection{Catheter Insertion Settings}

Participants aligned the virtual bed and mannequin with their physical workspace to match reach distance and ergonomics. Using the left controller, they grabbed the probe and adjusted bed height via raise, lower, and auto-align buttons; using the right controller, they grasped and advanced the needle, which remained active only while held. The system included a fixed monitor and movable tablet ultrasound viewer. During each session, it logged attempts, successes, probe alignment, and insertion time, with insertion time reported as the primary measure of procedural efficiency.

\subsection{Designing the Gamified Simulation}

\begin{figure}[t]
  \centering
    \includegraphics[width=1\linewidth]{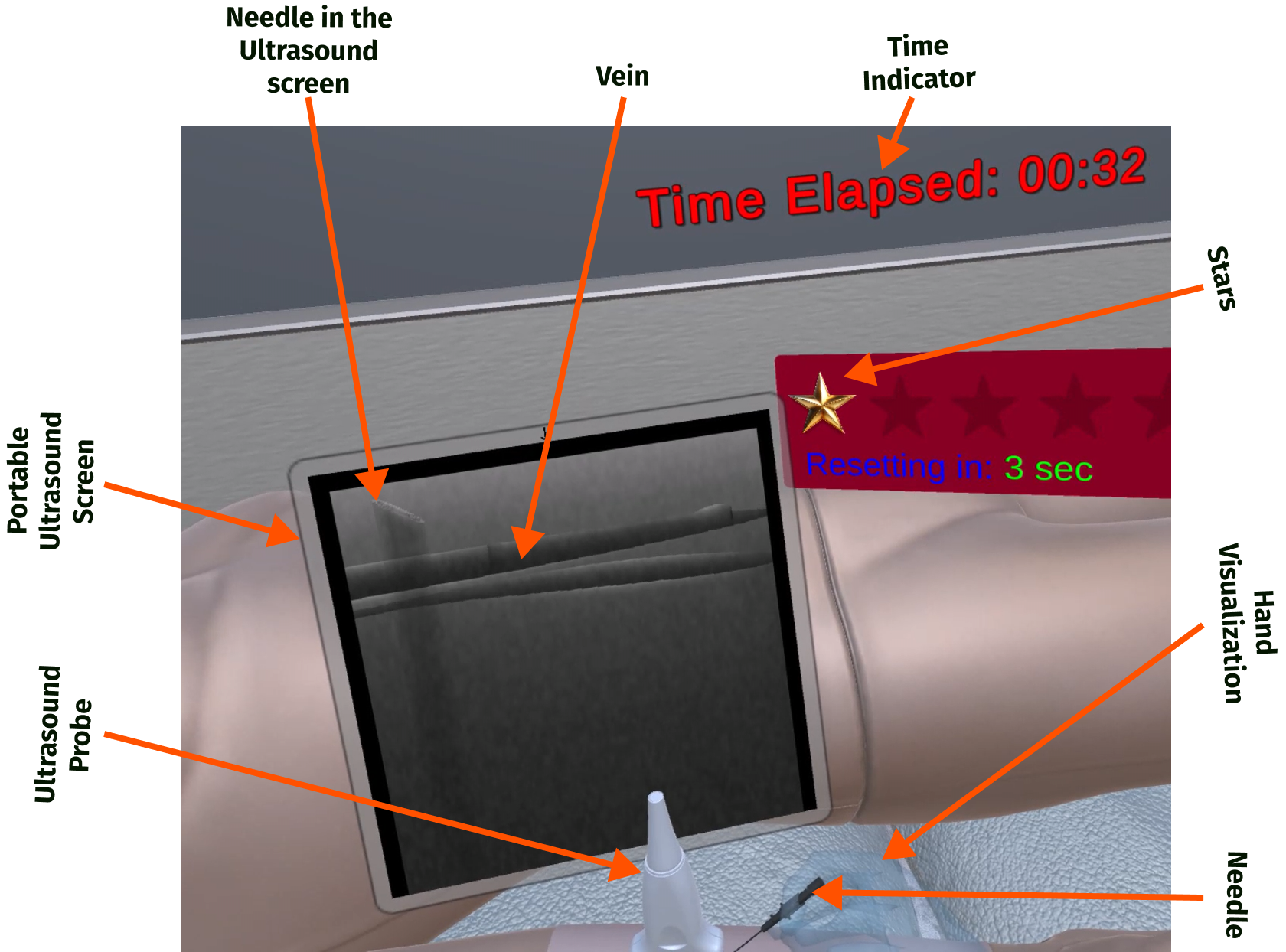}
  \caption{Gamified VR interface showing real-time ultrasound imaging, star-based feedback, needle and probe positions, and task progress indicators.}
  \label{fig:gamified_interface}
\end{figure}

\begin{figure}
    \centering
    \includegraphics[height=0.4\linewidth]{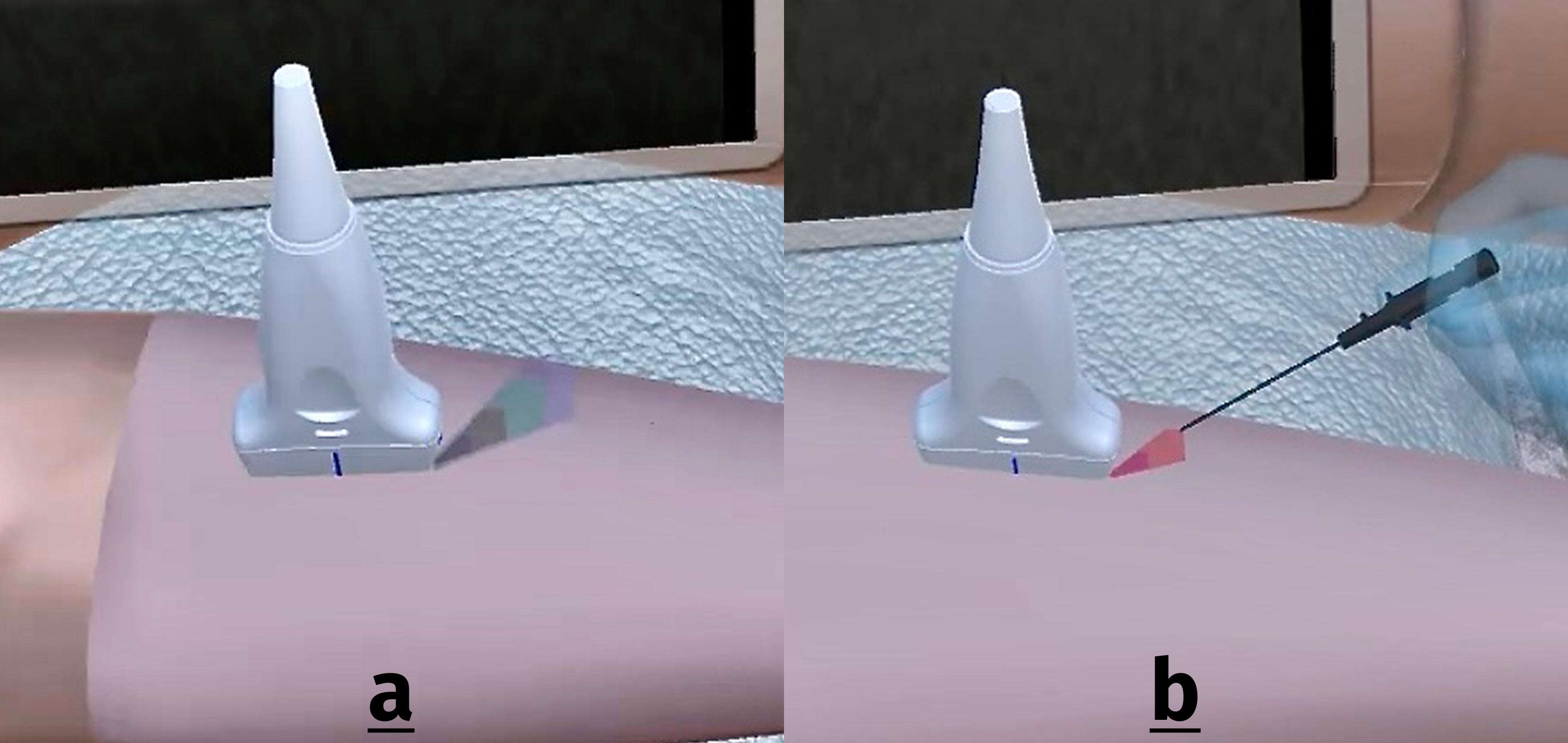}
    \caption{The landing-window alignment mechanism. (a) The triangular landing window appears when guidance is enabled, providing a target zone for correct needle approach. (b) As the needle advances through the window along the correct trajectory, the guide gradually disappears.
  }
    \label{fig:landing_window}
\end{figure}

The gamified version added structured progression and motivational feedback while preserving procedural fidelity. A cumulative progress system displayed five hollow stars per level; each successful insertion filled one star, briefly turning it gold for immediate reinforcement before it remained filled (Figure~\autoref{fig:gamified_interface}). To support motor precision, a \textbf{landing window} guidance mechanism~\cite{doesburg2025landing} displayed a semi-transparent circular guide near the target, changing from red to green based on alignment and fading as alignment improved (Figure~\autoref{fig:landing_window}). After each level, participants received a performance summary showing attempts, successes, completion time, and a tiered badge from \textit{Apprentice} to \textit{Grandmaster}. Subtle success tones complemented the visual feedback. All metrics were logged using the same architecture as the base simulation, enabling direct comparison across conditions.

After each level, participants viewed a \textbf{performance summary} showing attempts, successful insertions, and completion time, along with a tiered digital badge (e.g., \textit{Apprentice} to \textit{Grandmaster}) to support reflection and sustained motivation. Subtle success tones complemented the visuals. All metrics were logged using the same architecture as the base simulation to ensure direct comparability between conditions while maintaining medical authenticity.

\section{Motivation and Hypothesis}

Effective VR medical training requires procedural realism, engagement, structured feedback, and motivation. Although VR enables safe, repeatable practice for ultrasound-guided catheter insertion, learners may struggle to monitor progress and sustain focus. Gamification addresses these limitations through goal structures and feedback cues; therefore, we compared a standard simulation (\autoref{sec:design}) with a gamified variant.

Using a between-subjects design with one independent variable, \textit{Condition} (\textit{Standard} vs.\ \textit{Gamified}), we applied mixed methods, combining quantitative measures with post-session interviews. We examined effects on \textit{task performance}, \textit{usability perception}, and \textit{player experience} through two research questions.

\textbf{RQ1. Does gamification improve procedural efficiency in VR ultrasound-guided catheter insertion?} Since gamification can accelerate learning and motor coordination through immediate feedback and structured progression~\cite{pimentel2024gamification, merino2025enhancing}, we hypothesize: \textit{H1a:} Gamification improves task performance; \textit{H1b:} Gamification improves learning progression across levels.

\textbf{RQ2. How does gamification affect subjective experience?} Because rewards and feedback can improve usability, reduce workload, and support intrinsic motivation~\cite{sailer2017gamification, krath2021revealing, li2024gamification}, we hypothesize: \textit{H2a:} Gamification increases usability; \textit{H2b:} Gamification reduces perceived workload.

\section{Study--1: Gamification for Novice UG Catheter Insertion}

\subsection{Measures}

We used mixed quantitative and qualitative measures to assess performance, usability, workload, and experience. Objective performance was measured by \textbf{Insertion Time (IT)}, defined as the time from task initiation to successful insertion, reset after each success and averaged per participant and level. Subjective outcomes were assessed using the \textit{System Usability Scale (SUS)}~\cite{lewis2018system}, \textit{Player Experience Inventory (PXI)}~\cite{vanden2020development}, and \textit{NASA-TLX}~\cite{hart1988development}, supplemented by brief post-session interviews.

\begin{figure*}[t]
    \centering
    \includegraphics[width=0.9 \linewidth]{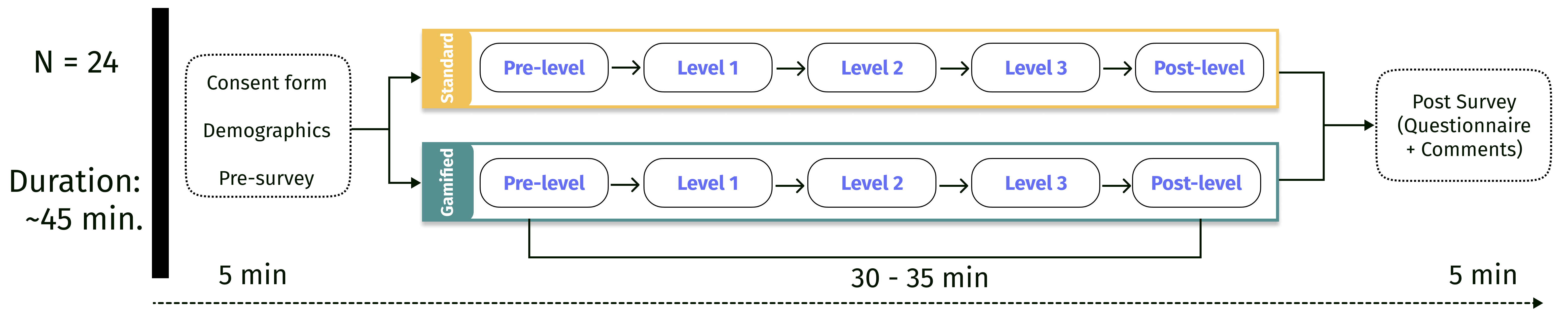}
   \caption{Overview of the Study procedure.}
    \label{fig:study_procedure}
\end{figure*}

\subsection{Participants}

An a priori power analysis ($d = 0.5$, power = 0.8, $\alpha = .05$)~\cite{faul2007g} indicated a minimum sample of 24 novices, recruited via snowball sampling. Participants were aged 20--40 years ($M$ = 26.38, $SD$ = 4.48), including 11 males and 13 females from diverse professional backgrounds. All were right-handed; 19 reported right-eye dominance. Twenty had normal vision and four had corrected-to-normal vision. VR experience varied: seven had none, six had used VR 1--5 times, and eleven more than five times. Sixteen played digital games at least monthly, and eight also played board games.

\subsection{Procedure}

Participants completed a 45-minute session in either the \textsc{Standard} or \textsc{Gamified} condition. After consent, briefing, a recorded VR catheter-insertion demo, and brief system familiarization, they completed a demographic questionnaire and proceeded to their assigned condition, with allocation balanced via a Latin Square (Figure~\ref{fig:study_procedure}). Participants were instructed to insert as quickly and accurately as possible; baseline and post-test tasks required success, with retries allowed. The study was approved by the institutional research ethics board.

\paragraph{Standard Condition.}
Participants completed ultrasound-guided insertions without gamified elements across five stages: a one-insertion \textit{pre-insertion} baseline, three practice levels, and a \textit{post-insertion} task identical to baseline. \textit{Level 1} required five successful insertions at a natural pace; \textit{Level 2} required five insertions emphasizing speed and accuracy; and \textit{Level 3} allowed five minutes to complete as many successful insertions as possible.

\paragraph{Gamified Condition.}
The sequence matched the \textsc{Standard} condition but added star rewards, auditory cues, landing-window guidance, performance summaries, and tiered badges. The \textit{pre-insertion} task remained ungamified. Across Levels 1--3, successful insertions filled stars, triggered feedback, and contributed to performance summaries and badges, with Level 2 emphasizing speed--accuracy and Level 3 using a five-minute performance window. The session ended with the ungamified \textit{post-insertion} task. Multimodal feedback was designed to support engagement and learning while preserving medical realism.

\subsection{Results}

Data were preprocessed and visualized in JMP and analyzed in SPSS 29.0. Normality was assessed using skewness and kurtosis $\pm1$; non-normal data were log-transformed.

\begin{figure}[hbt!]
    \centering

    \begin{subfigure}[t]{0.4\columnwidth}
        \includegraphics[height=.95\linewidth]{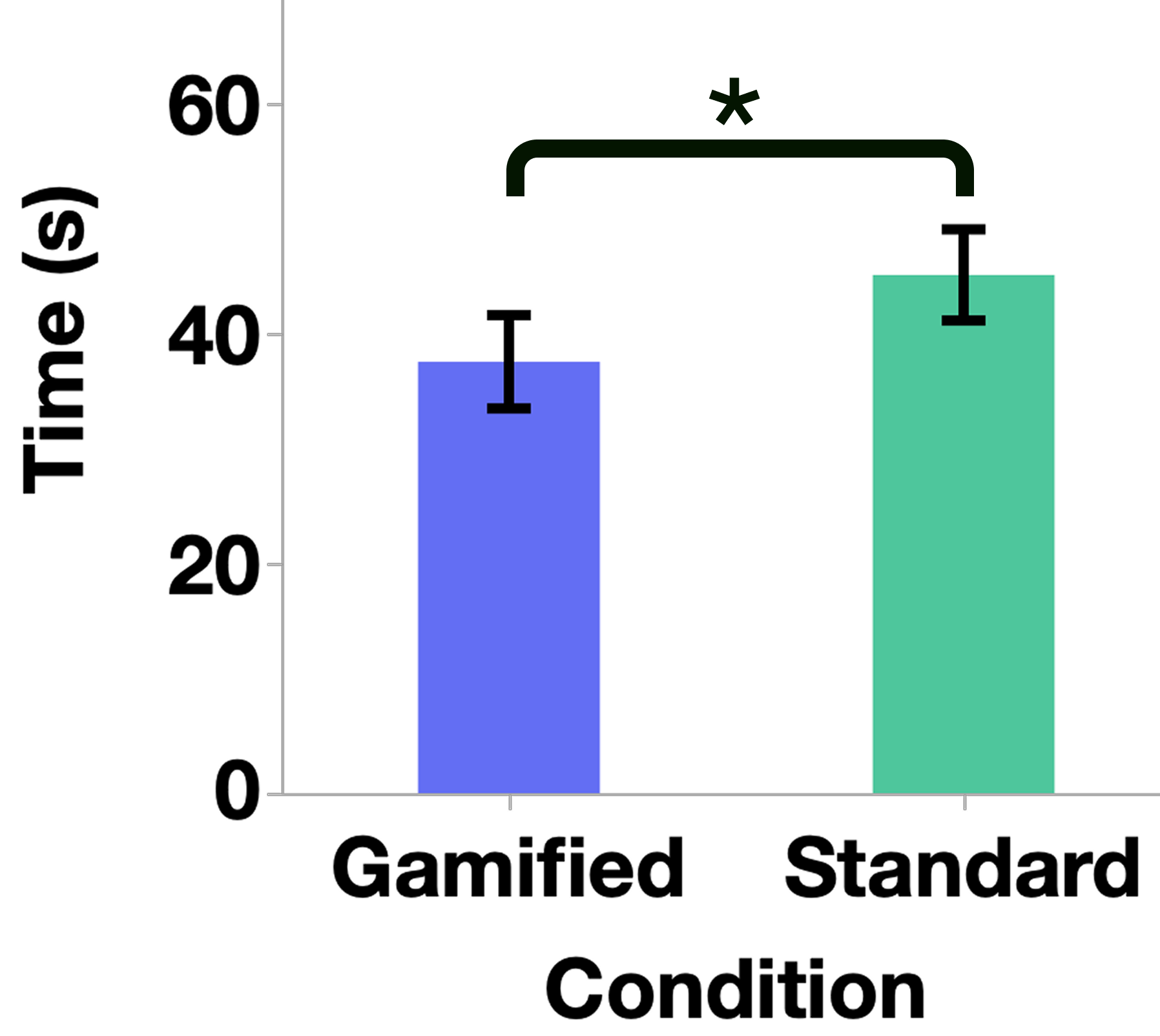}
        \caption{}\label{fig:time-non-expert}
    \end{subfigure}
    \hspace{0.05\textwidth}
    \begin{subfigure}[t]{0.4\columnwidth}
        \includegraphics[height=0.8\linewidth]{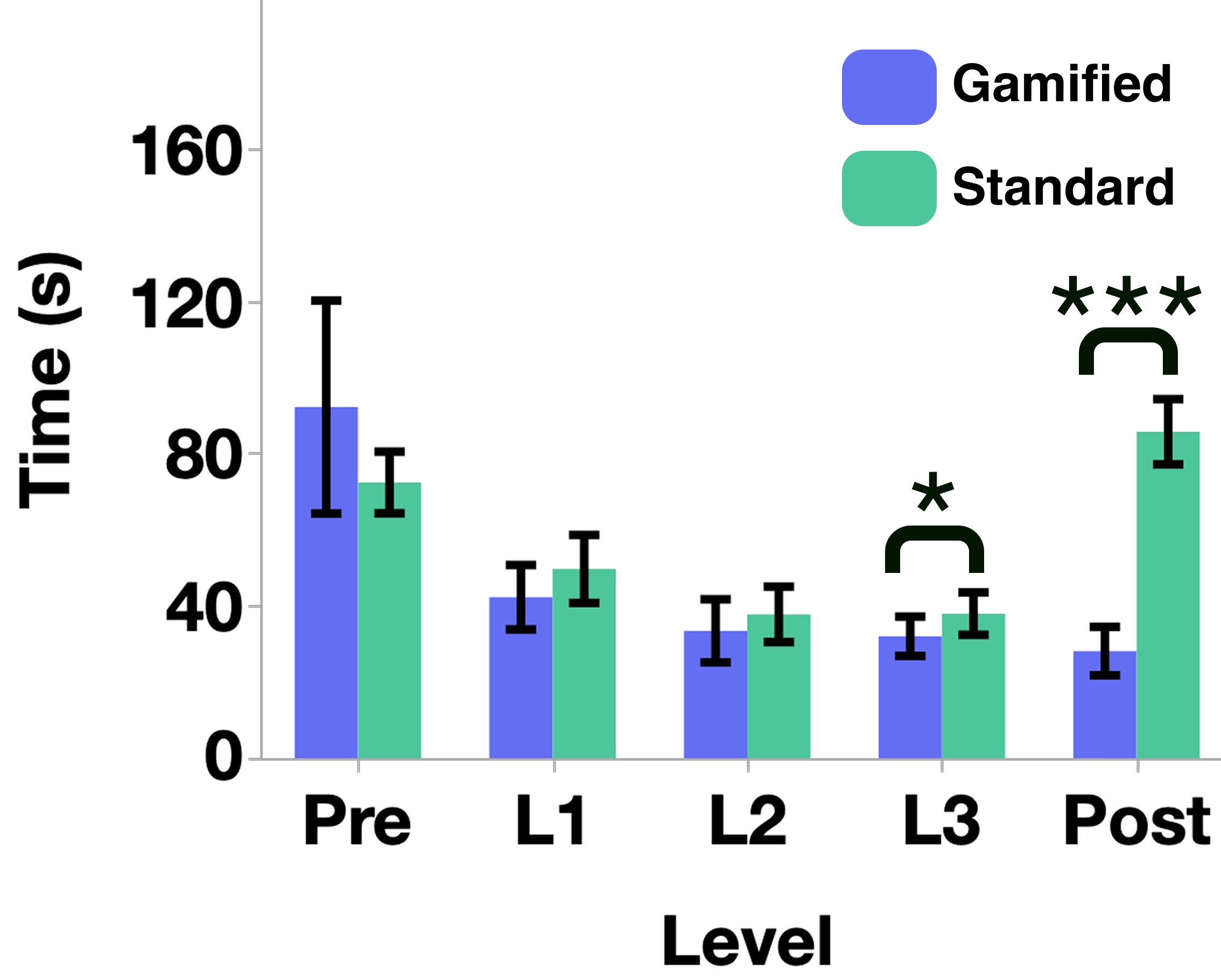}
        \caption{}\label{fig:time-level-non-expert}
    \end{subfigure}

    \caption{Insertion time results (in seconds) across conditions and levels. (a) IT by condition. (b) IT across levels.}  
    \label{fig:TCT}
\end{figure}

\textbf{IT:} A mixed ANOVA showed a significant main effect of Condition, $(F(1,22)=6.79,~p<.001,~\eta^2_{p}=.236)$, with the \textsc{Gamified} group completing insertions faster $(M=46.57,~SE=2.94)$ than the \textsc{Standard} group $(M=57.41,~SE=2.94)$ (\autoref{fig:time-non-expert}). A significant main effect of Level was also observed, $(F(4,88)=28.58,~p<.001,~\eta^2_{p}=.565)$, indicating learning over trials. The Condition $\times$ Level interaction was significant, $(F(4,88)=16.34,~p<.001,~\eta^2_{p}=.426)$, with group differences emerging at Level~3 $(p<.031)$ and the Post-level $(p<.001)$ but not earlier levels (\autoref{fig:time-level-non-expert}).

\begin{figure}
    \centering
    \includegraphics[width=1\linewidth]{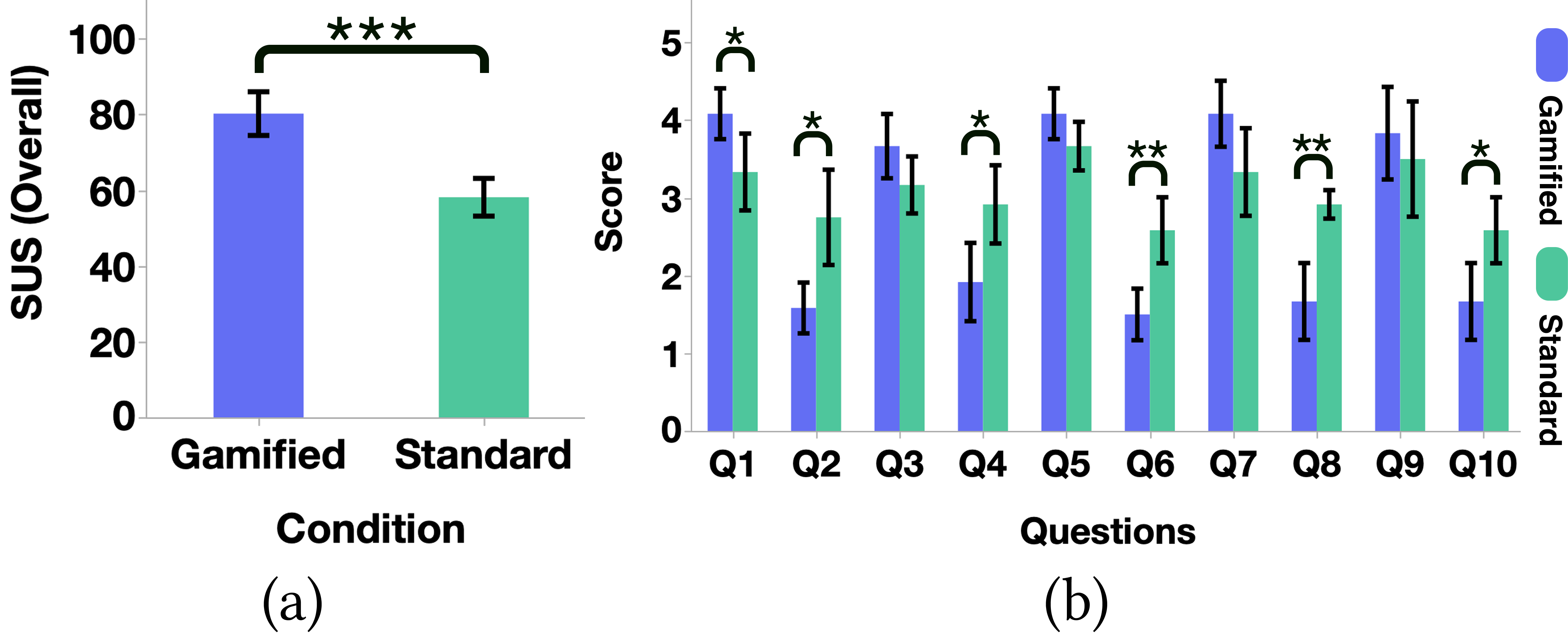}
    \caption{SUS scores (a) across conditions and (b) individual items.}
    \label{fig:SUS-S}
\end{figure}

\textbf{SUS:} A Friedman test revealed a significant effect of Condition, $(\chi^2(1)=11.00,~p<.001)$, with higher usability ratings in the \textsc{Gamified} condition (\autoref{fig:SUS-S}a). \autoref{tab:sus_s1} summarizes SUS scores and usability grades.

\begin{table}[h]
\centering
\caption{SUS Scores and Grades for two catheter insertion condition.}
\label{tab:sus_s1}
\begin{tabular}{|l|c|c|}
\hline
Condition           & Score & \multicolumn{1}{l|}{Grade} \\ \hline
\textbf{Gamified} & 78.54 & \textit{B}                 \\ \hline
\textbf{Standard}       & 58.13 & \textit{D}                 \\ \hline
\end{tabular}%
\end{table}

We observed significant differences across several SUS items between conditions (\autoref{fig:SUS-S}b). Participants preferred the \textsc{Gamified} condition for Q1 (frequency of use; $Z=-2.310,~p<0.021$). The \textsc{Standard} condition was rated as more complex (Q2; $Z=-2.565,~p<0.010$), requiring more technical support (Q4; $Z=-2.588,~p<0.010$), more inconsistent (Q6; $Z=-2.970,~p<0.003$), more cumbersome (Q8; $Z=-2.762,~p<0.006$), and requiring more learning before use (Q10; $Z=-2.373,~p<0.018$).

\begin{figure}
    \centering
    \includegraphics[width=.7\linewidth]{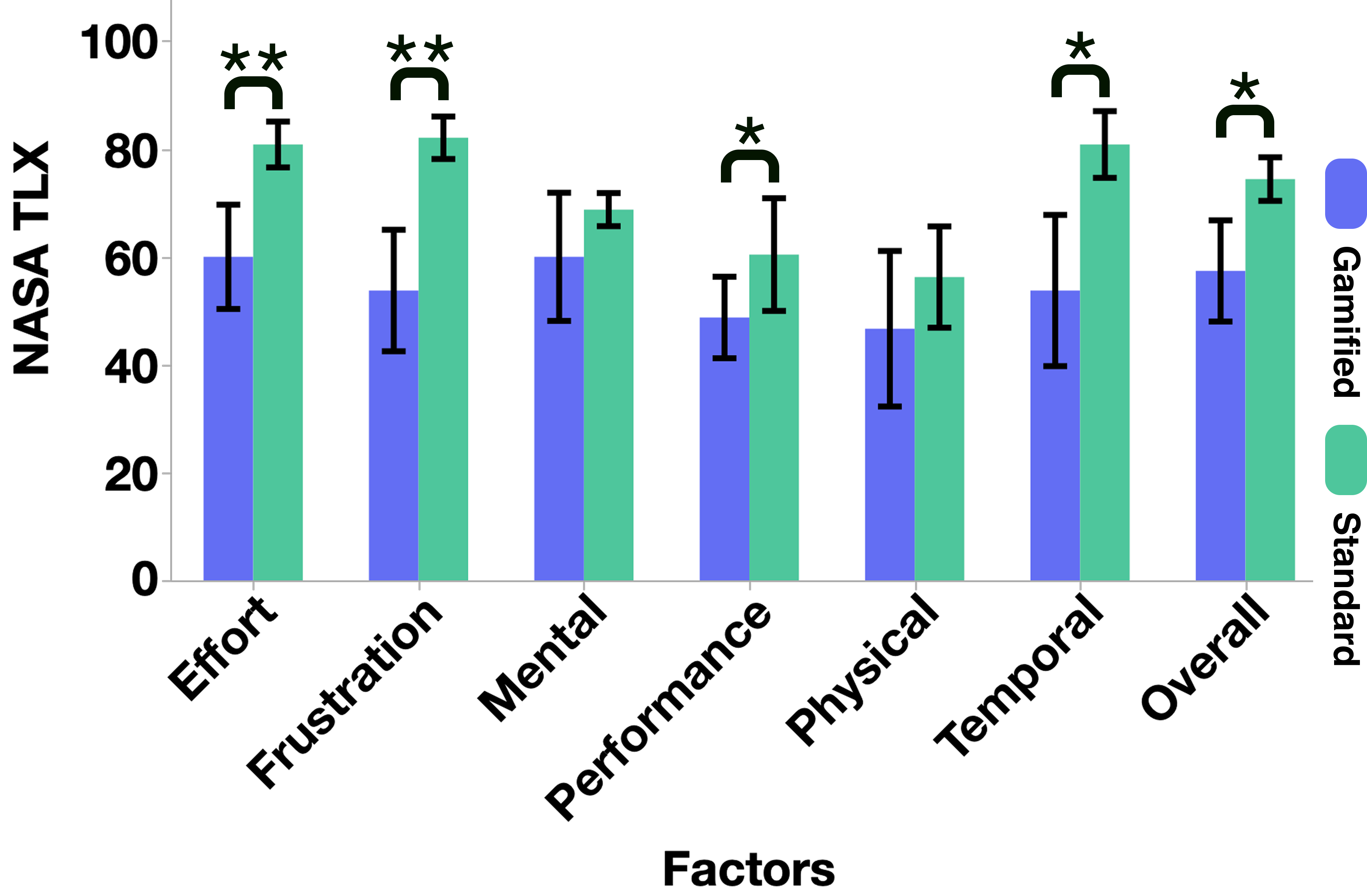}
    \caption{NASA TLX results commparing \textsc{Gamified} and \textsc{Standard} conditions.}
    \label{fig:nasa-s}
\end{figure}

\textbf{NASA TLX:} A Friedman test showed a significant overall workload difference between \textsc{Gamified} and \textsc{Standard} ($\chi^2(1)=4.455,~p<0.035$), with participants reporting lower workload in the \textsc{Gamified} condition (\autoref{fig:nasa-s}). Wilcoxon tests indicated significantly lower effort ($Z=-2.854,~p<0.004$), frustration ($Z=-2.809,~p<0.005$), and temporal demand ($Z=-2.555,~p<0.011$), alongside higher perceived performance ($Z=-2.149,~p<0.032$) in the \textsc{Gamified} condition. Mental and physical demand differences were not significant, suggesting comparable cognitive and physical requirements across systems.

\begin{figure}
    \centering
    \includegraphics[width=.85\linewidth]{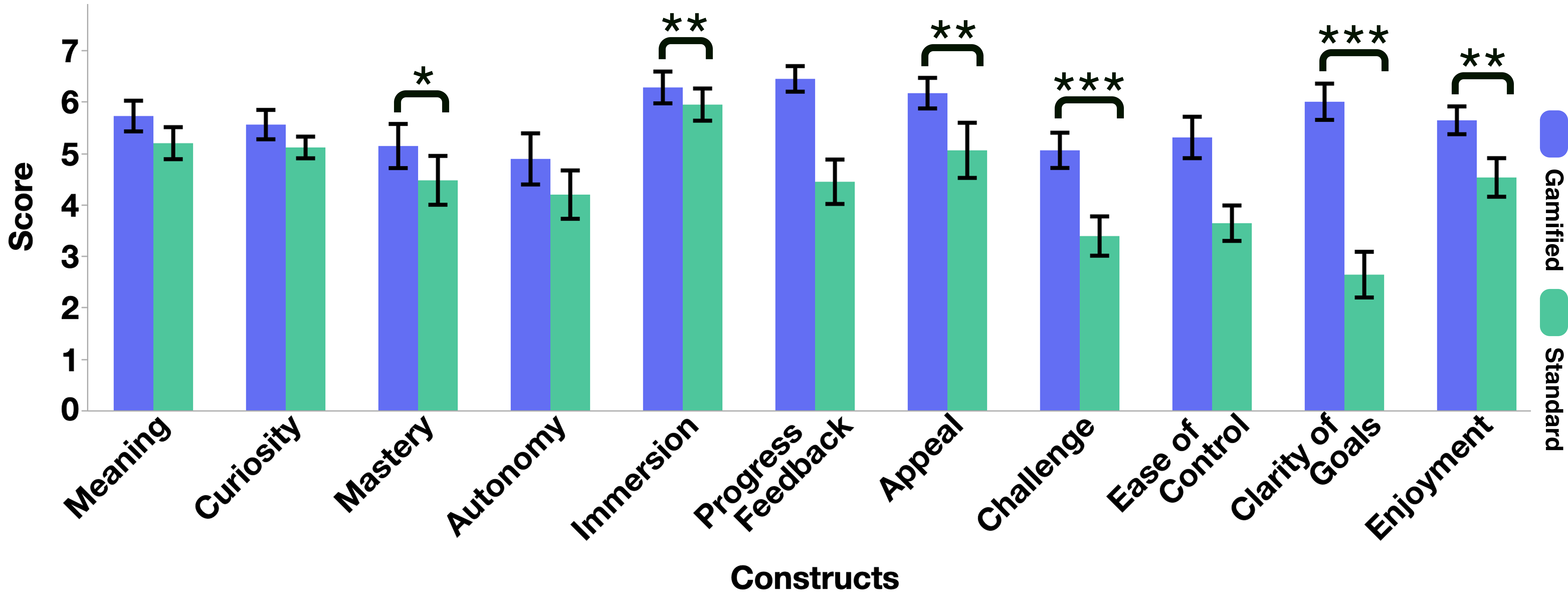}
    \caption{PXI results for \textsc{Gamified} and \textsc{Standard} conditions.}
    \label{fig:pxi-s}
\end{figure}

\textbf{PXI:} The 33-item PXI was analyzed using independent-samples t-tests between \textsc{Gamified} and \textsc{Standard}. Significant effects favored the \textsc{Gamified} condition (\autoref{fig:pxi-s}), including higher appeal $(t(22)=2.97,~p<.008)$, challenge $(t(22)=4.52,~p<.001)$, clarity of goals $(t(22)=7.98,~p<.001)$, enjoyment $(t(22)=3.12,~p<.005)$, immersion $(t(22)=3.45,~p<.002)$, and mastery $(t(22)=2.79,~p<.010)$. Differences in autonomy, curiosity, and relatedness were not significant after Holm correction. Overall, gamification improved experiential quality—enhancing engagement, competence, goal clarity, and enjoyment—whereas the \textsc{Standard} condition showed weaker immersion and feedback perception.

\textbf{Interview:} \textsc{Gamified} participants reported greater engagement, motivation, and performance awareness. They found the game elements enjoyable and informative (P2), with real-time cues supporting progress tracking and success recognition (P5). Visible scores and progression improved goal clarity and motivation (P9), encouraged focus on improvement across levels (P11), and increased appreciation of the procedure's real-world difficulty (P12). In contrast, \textsc{Standard} participants reported uncertainty, difficulty, and lower engagement, citing repeated attempts without clear feedback (P1), difficulty interpreting ultrasound views (P4), uncertainty about correct execution (P5), need for more guidance and alerts (P6), and frustration when progress stalled (P9).

\subsection{Discussion}

Study~1 shows that gamification benefits novices by improving task speed, usability, workload, engagement, and goal clarity. These results suggest that task-aligned gamification supports early skill acquisition in ultrasound-guided catheter insertion. To assess whether these benefits generalize beyond novices, we conducted Study~2 with experienced clinicians.

\section{Study--2: Gamification on Experts' UG Catheter Insertion}

\subsection{Overview and Motivation}

Study~2 examined whether gamification benefits expert clinicians, whose established procedural knowledge, perceptual--motor skills, and mental models may influence how they respond to feedback and motivational cues. We therefore evaluated whether gamification affects performance, workload, and user experience when baseline proficiency is high.

\subsection{Participants}

An a priori power analysis (Cohen's $d=0.5$, power = 0.8, $\alpha=0.05$)~\cite{faul2007g} indicated a minimum sample of 12. We recruited 12 experts through professional networks and clinician invitations. Participants were aged 25--63 years ($M=37.08$, $SD=11.16$), including 10 males and 2 females from diverse medical and technical backgrounds. All were right-handed; ten reported right-eye dominance and two left-eye dominance. Seven had normal vision and five had corrected-to-normal vision, with no color or binocular impairments. Most had no prior VR experience (9), two had used VR 1--5 times, one had used it more than five times, and none reported regular digital gameplay.

\subsection{Results}

\begin{figure}[hbt!]
    \centering

    \begin{subfigure}[t]{0.4\columnwidth}
        \includegraphics[height=.95\linewidth]{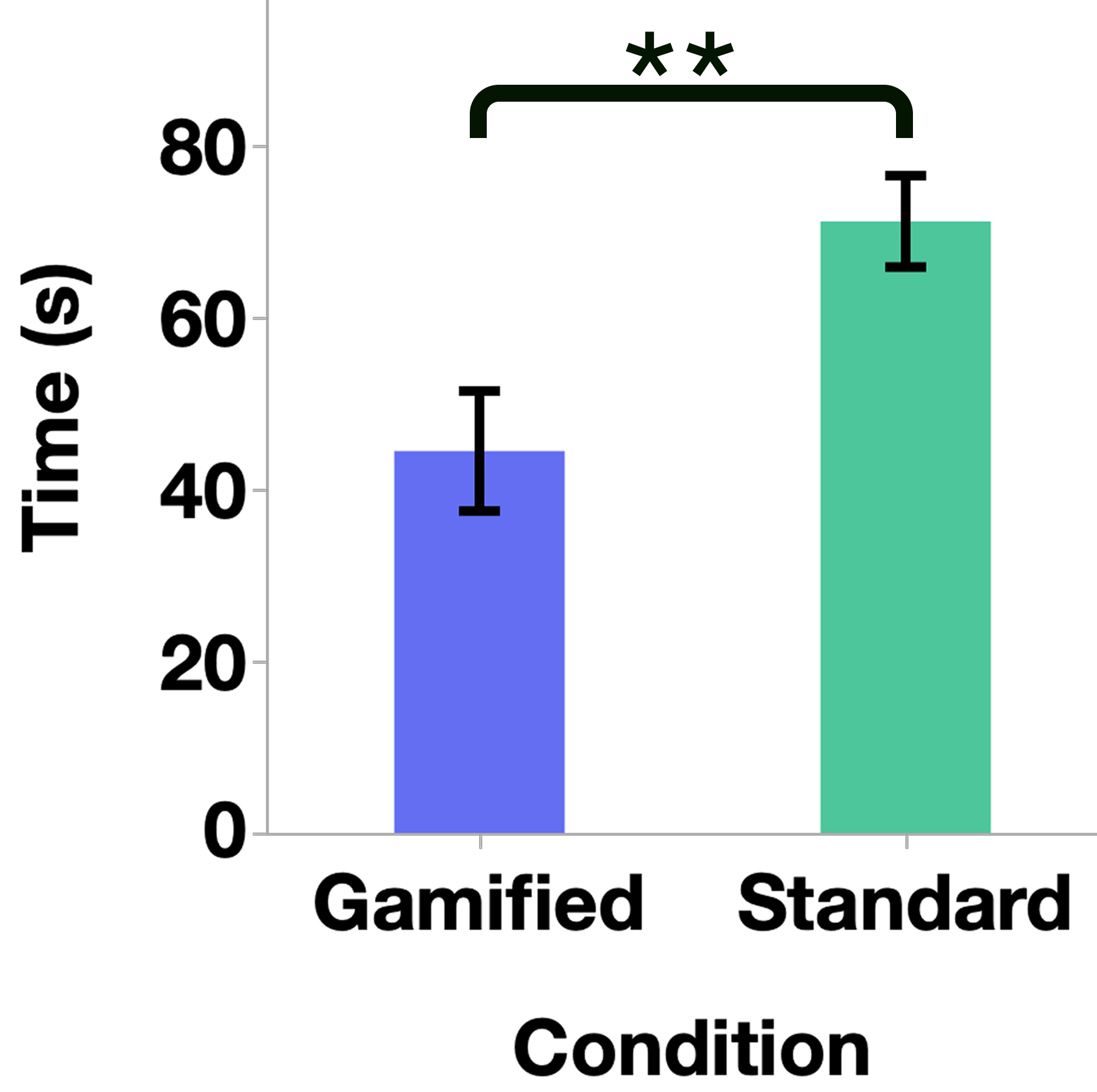}
        \caption{}\label{fig:time-expert}
    \end{subfigure}
    \hspace{0.05\textwidth}
    \begin{subfigure}[t]{0.4\columnwidth}
        \includegraphics[height=0.8\linewidth]{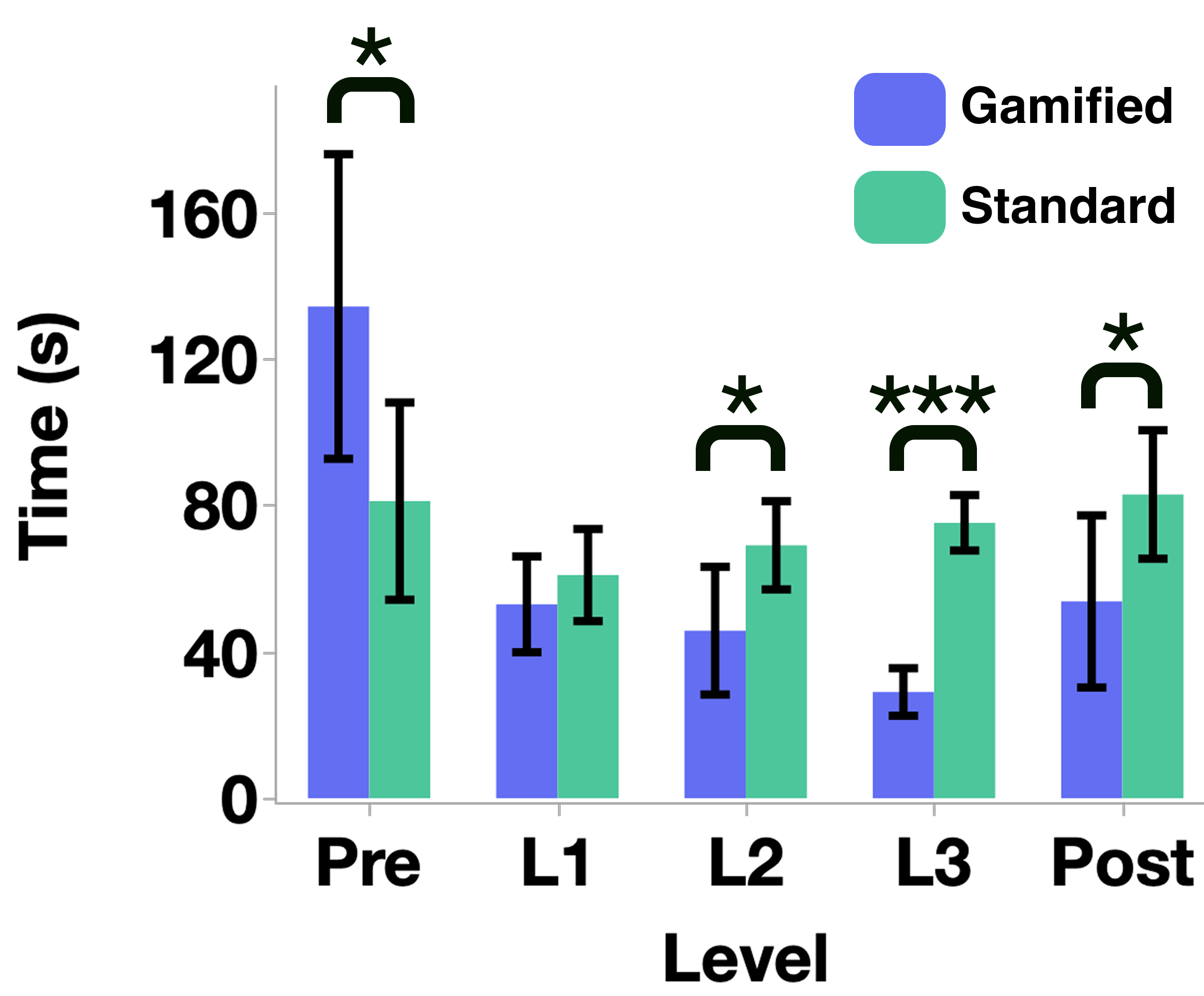}
        \caption{}\label{fig:time-level-expert}
    \end{subfigure}
    \caption{IT results for expert participants. (a) Mean IT by condition. (b) IT across levels.}  
    \label{fig:TCT-E}
\end{figure}

\textbf{IT:} A mixed ANOVA showed a significant main effect of Condition, $(F(1,10)=11.21,~p<.007,~\eta^2_{p}=.528)$, with the \textsc{Gamified} group completing tasks faster $(M=45)$ than the \textsc{Standard} group $(M=75)$ (\autoref{fig:time-expert}), indicating improved procedural fluency even among experts. A significant main effect of Level was also observed, $(F(2.97,29.74)=17.79,~p<.001,~\eta^2=.640)$, reflecting progressive learning; Bonferroni tests showed post-level performance $(M=4.60)$ was significantly faster than Levels~1–3 $(p<.001)$.

The Condition $\times$ Level interaction was significant, $(F(2.97,29.74)=11.48,~p<.001,~\eta^2=.534)$, indicating different improvement rates across conditions (\autoref{fig:time-level-expert}). Both groups improved, but the \textsc{Gamified} group showed faster gains and consistently lower ITs. In\textsc{Gamified} condition, we observed significant difference for  L2 $(p<.022)$, L3 $(p<.001)$, L4 $(p<.023)$, and Post $(p<.022)$, indicating gamification benefits in later stages.

\begin{figure}
    \centering
    \includegraphics[width=1\linewidth]{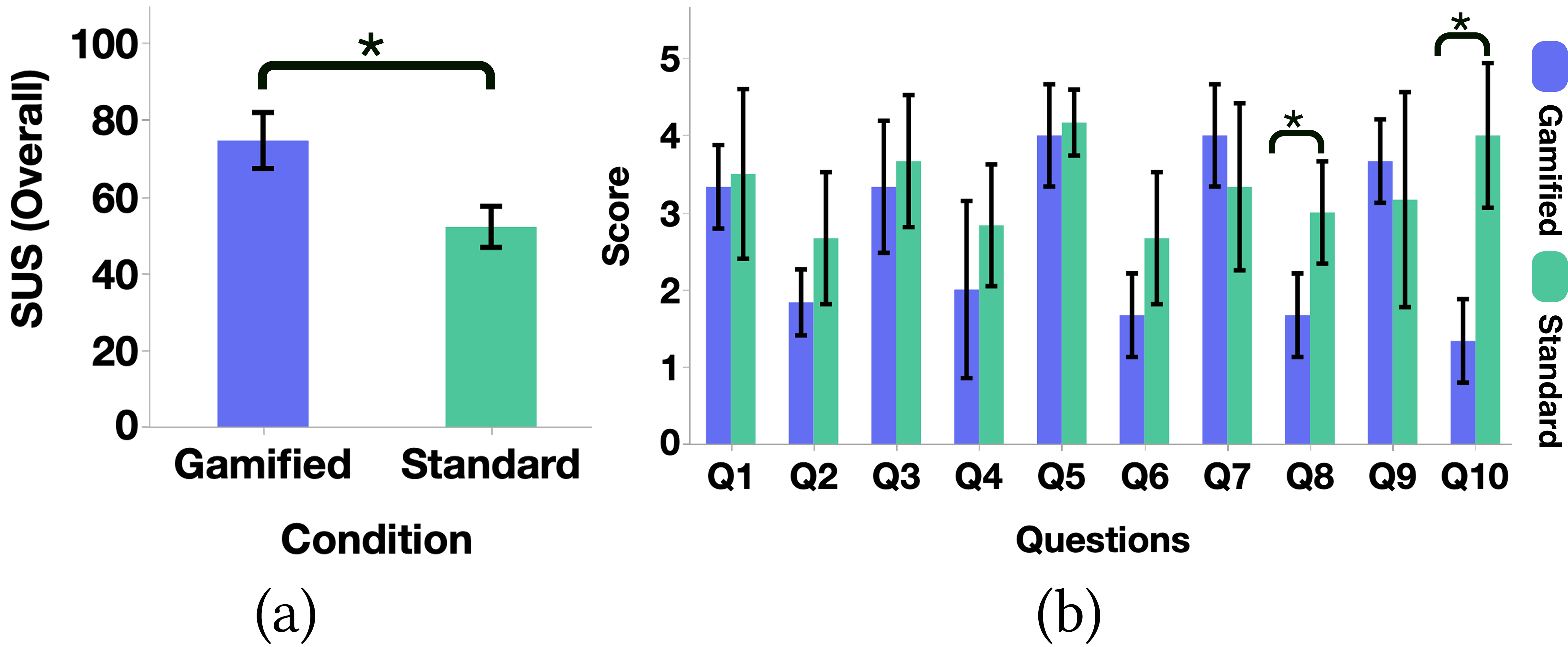}
    \caption{SUS results for expert participants. (a) Overall scores show higher usability for the \textsc{Gamified} condition. (b) Item-wise ratings (Q1–Q10).}
    \label{fig:SUS-E}
\end{figure}

\textbf{SUS:} A Friedman test revealed a significant usability difference ($\chi^2(1)=6.00,~p<.014$), with experts rating the \textsc{Gamified} system higher. The \textsc{Gamified} condition achieved a mean SUS of 74 (``Good'') versus 54 (``Poor'') for \textsc{Standard} (\autoref{fig:SUS-E}a, \autoref{tab:sus_s2}).

\begin{table}[h]
\centering
\caption{SUS Scores and Grades for two catheter insertion conditions.}
\label{tab:sus_s2}
\begin{tabular}{|l|c|c|}
\hline
Condition           & Score & \multicolumn{1}{l|}{Grade} \\ \hline
\textbf{Gamified} & 74.58 & \textit{B}                 \\ \hline
\textbf{Standard}       & 52.08 & \textit{D}                 \\ \hline
\end{tabular}%
\end{table}

Wilcoxon Signed-Rank tests on SUS items showed significant differences for Q8 (“cumbersome”; $Z=-2.27,~p<.023$) and Q10 (“needed to learn a lot”; $Z=-2.23,~p<.026$), both favoring the \textsc{Gamified} condition (\autoref{fig:SUS-E}b). While other items were not significant, the overall trend favored gamification, indicating improved usability and learnability even for experts.

\begin{figure}
    \centering
    \includegraphics[width=.5\linewidth]{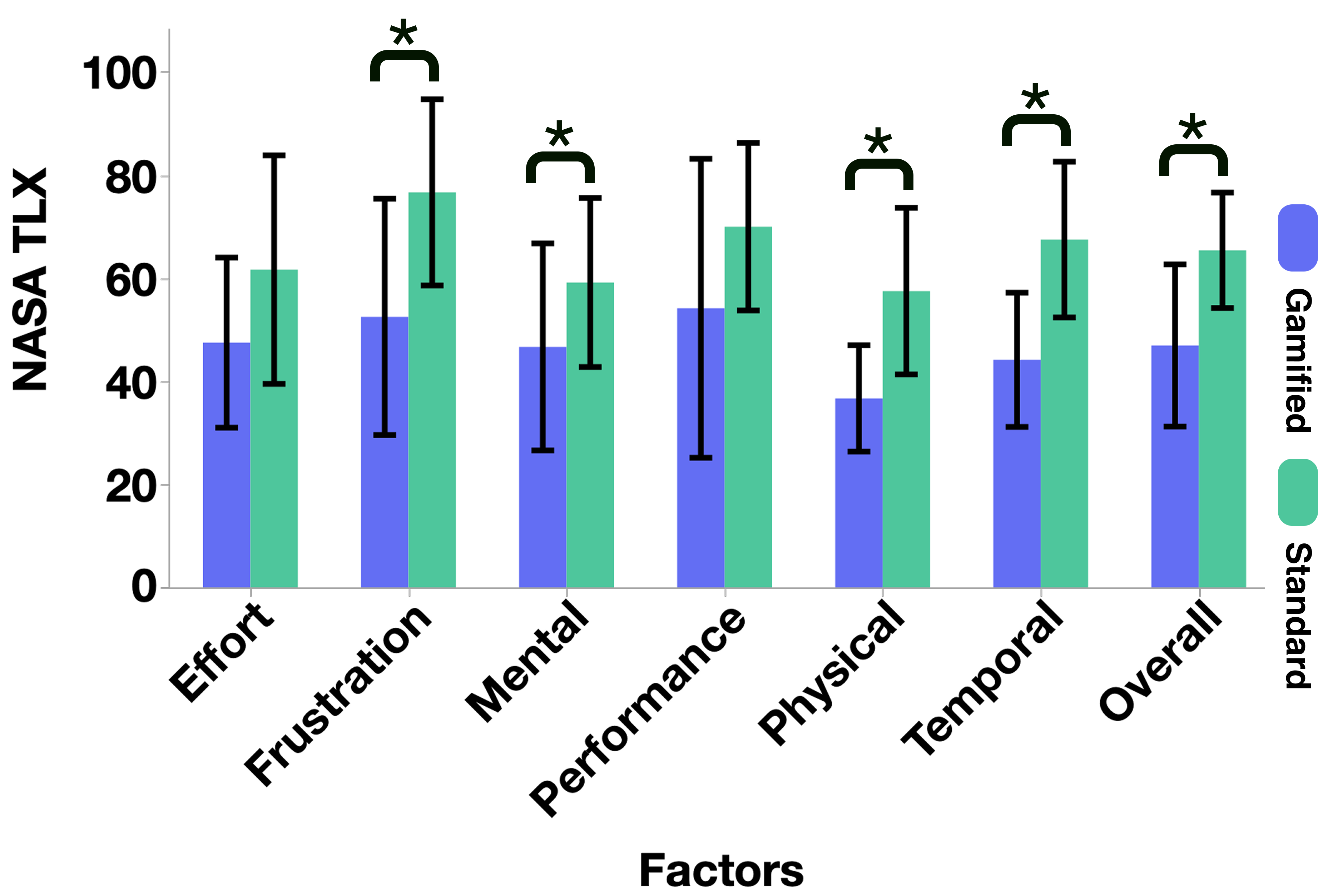}
    \caption{NASA TLX results comparing \textsc{Gamified} and \textsc{Standard} conditions for expert participants.}
    \label{fig:nasa-e}
\end{figure}

\textbf{NASA-TLX:} A Friedman test showed a significant workload difference between conditions $(\chi^2(1)=6.00,~p<.014)$, with experts reporting lower workload in the \textsc{Gamified} condition (\autoref{fig:nasa-e}). Mean scores were lower across most dimensions, suggesting game-based feedback reduced cognitive and physical strain.

Wilcoxon tests identified significant reductions in Frustration $(Z=-2.02,~p<.043)$, Mental Demand $(Z=-2.04,~p<.041)$, Physical Demand $(Z=-2.06,~p<.039)$, and Temporal Demand $(Z=-1.99,~p<.046)$ for the \textsc{Gamified} condition. Overall, gamification produced a more balanced workload profile, indicating that even experts benefited from structured feedback and progression cues during repetitive procedural tasks.

\begin{figure}
    \centering
    \includegraphics[width=.8\linewidth]{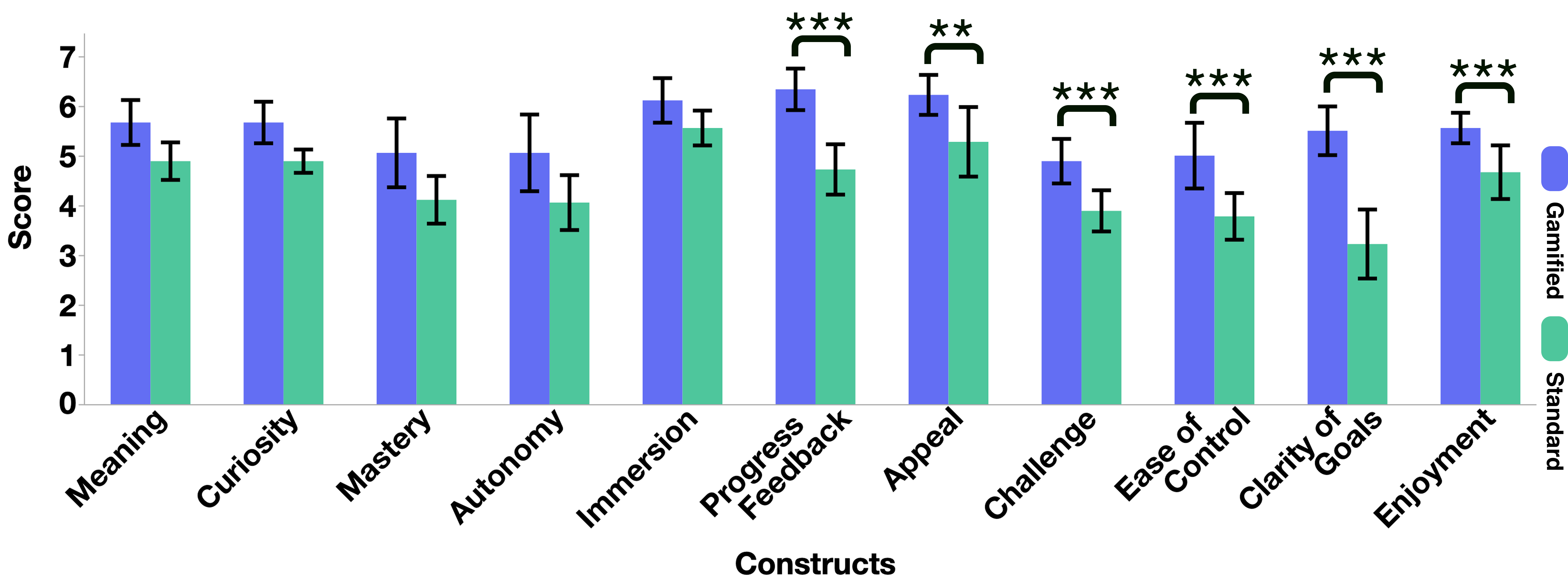}
    \caption{PXI results comparing \textsc{Gamified} and \textsc{Standard} conditions for expert participants.}
    \label{fig:pxi-e}
\end{figure}

\textbf{PXI:} Results indicate that gamification improved player experience, particularly on functional dimensions. Compared to \textsc{Standard}, the \textsc{Gamified} condition showed significantly higher Clarity of Goals $(t(10)=3.927,~p<.001)$, Challenge $(t(10)=2.577,~p<.028)$, Appeal $(t(10)=2.966,~p<.008)$, Progress Feedback $(t(10)=3.191,~p<.001)$, and Enjoyment $(t(10)=3.833,~p<.001)$. No psychological consequence measures reached significance.

Experts in the \textsc{Gamified} condition reported stronger control, clearer objectives, and more rewarding feedback, whereas participants reported lower motivation and structure in \textsc{Standard} condition. The largest gains occurred in Clarity of Goals $(\Delta M=2.28)$ and Progress Feedback $(\Delta M=1.61)$, highlighting the role of feedback and achievement cues.

\textbf{Interview:} Expert participants recognized the system’s training value despite differences from real practice. Some noted realism limits, such as fixed needle paths, while still finding the gamified version engaging and useful (P2, P3). Visible progress cues also reduced stress by making feedback explicit (P6). In contrast, \textsc{Standard} participants were more critical, citing oversensitivity, reduced realism, unclear targets, repeated failed insertions, and lower engagement (P1, P2, P5). Overall, gamification improved expert experience by strengthening goal awareness, perceived competence, and engagement.

\section{Discussion}

This work examined how visual cues, auditory reinforcement, and level-based progression affect VR ultrasound-guided PIVC training. Across novices and experts, gamification improved procedural efficiency, reduced workload, and enhanced usability and experience. Unlike prior VR medical training focused on realism or instructor-led debriefing~\cite{seymour2002virtual, mcgaghie2010critical, lateef2010simulation, alaker2016virtual}, our results show that \textit{continuous formative feedback} can make progress and alignment explicit, supporting self-regulated learning.

\subsection{Effects on Novice Training and Early Skill Acquisition}

Study~1 showed that novices in the \textsc{Gamified} condition completed insertions faster, especially in later stages, supporting \textbf{H1a} and \textbf{H1b}. Structured feedback through stars, alignment cues, and summaries accelerated spatial understanding and procedural sequencing. Subjective results showed higher usability and lower effort, frustration, and temporal demand. Consistent with Flow theory~\cite{csikszentmihalyi1997flow}, clear goals and immediate feedback reduced uncertainty and cognitive load. PXI gains in clarity, challenge, immersion, and mastery suggest that gamification served as instructional scaffolding, supporting competence and mental-model formation. These findings support \textbf{H2a} and \textbf{H2b}.

\subsection{Effects on Expert Performance and Skill Reinforcement}

Study~2 showed that gamification also benefited experts, improving completion times and reducing mental, physical, temporal demand, and frustration. For experts, gamification did not teach procedural knowledge but optimized execution by reducing self-monitoring demands and stabilizing pacing. PXI gains in goal clarity, challenge, enjoyment, feedback, and appeal show that structured progress cues remain useful during repetitive expert tasks. Longer times in the \textsc{Standard} expert group likely reflected lower VR familiarity and real-world habits that transferred poorly to VR. In contrast, gamified feedback reduced interpretation effort and improved efficiency, indicating benefits beyond gaming or VR familiarity.

\subsection{Cross-Study Synthesis}

Gamification supported both groups through different mechanisms: for novices, it acted as \textit{instructional scaffolding}; for experts, as \textit{cognitive optimization}. Importantly, benefits did not compromise clinical authenticity, a common concern in medical gamification~\cite{deterding2011game, nacke2017maturing, sailer2017gamification}. When aligned with task goals, elements such as alignment guides and progress indicators enhanced procedural engagement rather than distraction. Overall, real-time, perceptually salient feedback improved early skill acquisition and expert performance consistency.

\section{Limitations and Future Work}

This study used lightweight instructional gamification, limiting generalizability to richer designs~\cite{deterding2011game, hamari2014does, seaborn2015gamification}. The fixed single-arm anatomy reduced ecological validity, and only immediate performance was assessed, with a modest expert sample and no long-term learning measures. Future work should explore adaptive or fading feedback, richer gamification, anatomical variability, larger cohorts, longitudinal transfer, and improved haptics. As gamified cues are absent in clinical practice, reflection, debriefing, and fading feedback may help reduce reliance and negative transfer.

\section{Conclusion}

We evaluated a gamified VR ultrasound-guided catheter insertion simulator with novices and clinicians. Goal-aligned visual and auditory feedback improved efficiency, reduced workload, and enhanced usability and experience without compromising realism. Rather than acting as superficial motivation, gamification served as formative feedback that supported novice learning and expert fluency, offering design guidance for future catheter-insertion training systems.

\end{document}